\documentclass[twocolumn]{aastex63}
\usepackage{xspace,amsmath,amssymb,lineno,multirow}

\shorttitle{Metallicities of Five $z>5$ Emission-Line Galaxies in SMACS 0723 Revealed by JWST}
\shortauthors{Taylor et al.}

\begin{document}

\title{Metallicities of Five $z>5$ Emission-Line Galaxies in SMACS 0723 Revealed by JWST}

\correspondingauthor{A.~J.~Taylor}
\email{ataylor@astro.wisc.edu}

\author[0000-0003-1282-7454]{A.~J.~Taylor}
\affiliation{Department of Astronomy, University of Wisconsin--Madison,
475 N. Charter Street, Madison, WI 53706, USA}

\author[0000-0002-3306-1606]{A.~J.~Barger}
\affiliation{Department of Astronomy, University of Wisconsin--Madison,
475 N. Charter Street, Madison, WI 53706, USA}
\affiliation{Department of Physics and Astronomy, University of Hawaii,
2505 Correa Road, Honolulu, HI 96822, USA}
\affiliation{Institute for Astronomy, University of Hawaii, 2680 Woodlawn Drive,
Honolulu, HI 96822, USA}

\author[0000-0002-6319-1575]{L.~L.~Cowie}
\affiliation{Institute for Astronomy, University of Hawaii,
2680 Woodlawn Drive, Honolulu, HI 96822, USA}

\begin{abstract}
JWST's Early Release Observations of the lensing cluster SMACS J0723.3-7327 have given an unprecedented spectroscopic look into the high-redshift universe. These observations reveal five galaxies at $z>5$. All five have detectable [OIII]$\lambda$4363 line emission, indicating that these galaxies have high temperatures and low metallicities and that they are highly star forming. In recent work, the metallicities of these five galaxies have been studied using various techniques. Here we summarize and compare these previous results, as well as perform our own measurements of the metallicities using improved methodologies that optimize the extraction of the emission lines. In particular, we use simultaneous line fitting and a fixed Balmer decrement correction, as well as a novel footprint measurement of the emission lines in the 2D spectra, to produce higher fidelity line ratios that are less sensitive to calibration and systematic effects. We then compare our metallicities to those of $z\lesssim1$ galaxies with high rest-frame equivalent widths of H$\beta$, finding that they may be good analogs. Finally, we estimate that the JWST galaxies out to $z\sim8$ are young compared to the age of the universe.
\end{abstract}

\section{Introduction} \label{sec:intro}

Prior to the launch of the James Webb Space Telescope (JWST), the rest-frame optical high-redshift universe was mostly inaccessible. While instruments such as the MOSFIRE spectrograph on the Keck~II 10~m telescope can observe out to 2.4~$\mu$m, the abundance of sky lines and the diminished atmospheric transmission of 2.5--3.4~$\mu$m light have placed severe limits on our ability to observe emission-line galaxies beyond redshifts of $z\sim3$--4. The JWST Early Release Observations (ERO) of the lensing cluster SMACS J0723.3-7327 have definitively demonstrated that these limitations no longer exist. 

While previous spectroscopic studies of galaxies at $z>5$ were primarily limited to observations of the Ly$\alpha$ line \citep[e.g.][]{hu16, matthee15, santos16, jiang17, konno18, shibuya18, songaila18, songaila22, LAGER19, taylor20, taylor21, ning22, wold22}, the NIRSpec instrument on JWST now permits observations of rest-frame optical emission lines, such as [OIII]$\lambda\lambda\lambda$5007,4959,4363 and the Hydrogen Balmer series out to $z\sim9$. Of these lines, the [OIII]$\lambda$4363 auroral line is of particular interest. This line is typically only seen in high-temperature, low-metallicity galaxies, and it serves as an excellent electron temperature ($T_e$) diagnostic when compared to the line strength of the [OIII]$\lambda\lambda$5007,4959 complex \citep[e.g.,][]{pilyugin05,izotov06,yin07}. This ``direct $T_e$ method" has been used to determine gas phase metallicities of galaxies from the local universe to $z\sim3$ \citep[e.g.,][]{kakazu07,brown16,ly16,indah21,laseter22}. With the JWST SMACS J0723 observations, metallicities have now been measured by various authors to $z\sim8.5$.

\cite{carnall22} introduced the JWST/NIRSpec sample and their spectroscopic redshifts. They used the \texttt{Pandora.ez} tool \citep{garilli10} and visual inspection to determine redshifts for the 35 galaxies targeted by the microshutter array observations. They found secure redshifts for 10 galaxies, of which 5 (04590, 05144, 06355, 08140, and 10612) lie at $z>5$. They also noted that several of the spectra showed clear detections of the [OIII]$\lambda$4363 emission line.

\cite{schaerer22} used the level~3 1D spectra of objects 04590, 06355, and 10612 from the JWST Science Calibration Pipeline (version 1.5.3) as their dataset for measuring metallicities. 
For each galaxy, they averaged the 1D spectra from each pointing together after masking spectral regions 
that were affected by cosmic rays and other artifacts. They fit Gaussian profiles to the individual emission lines in these averaged 1D spectra after assuming a flat continuum in $f_{\lambda}$ to measure line fluxes for [OIII]$\lambda\lambda\lambda$5007,4959,4363, [OII]$\lambda$3727, [NeIII]$\lambda$3869, H$\beta$, H$\gamma$, and H$\delta$.  They noted that the Balmer line ratios for some of the objects were non-physical, featuring ratios larger than those predicted by Case B recombination \citep{osterbrock89}. To compensate for this, they applied a power-law correction to the spectral flux fit to the Case B ratios for H$\gamma$/H$\beta$ and H$\delta$/H$\beta$. This correction loses the information on any dust attenuation. Using their corrected line fluxes, they followed \cite{izotov06} to derive direct $T_e$ method metallicities for their three objects. They found an unreasonably high value of $T_e$ for galaxy 04590, so they did not publish it.

\cite{curti22} also calculated metallicities for objects 04590, 06355, and 10612. 
However, they retrieved level 2 2D spectral data products from the JWST Science Calibration Pipeline (version 1.5.3) and reprocessed them using the GTO Pipeline (NIRSpec GTO collaboration, in prep) with optimized extraction apertures and bad/cosmic ray pixel masking. They also used a response function calibration based on the JWST observed calibration star 2MASS J18083474+6927286 (JWST Program \#1128). Through this reprocessing, they noted that one of the microshutters in the array failed to open for one of the nods in observation 7 of object 04950. They used the GTO Pipeline to combine the spectra obtained in each of the two observations, excluding the failed nod for 04950. They made their resulting 1D spectra publicly available\footnote{\url{https://doi.org/10.5281/zenodo.6940561}}. 

\cite{curti22} used \texttt{PPXF} \citep{cappellari17} to fit both lines fluxes and continua measurements to the spectra. They found that their Balmer ratios also showed deviations from Case B recombination, but only in H$\delta$ and higher energy lines. They followed \cite{nicholls13} to determine $T_e$ and the \texttt{getIonAbundance} routine from \texttt{PYNEB} \citep{luridiana12,luridiana15} to calculate abundances of O$^+$ and O$^{++}$ in each galaxy. They also noted a high value of $T_e$ (27700~K) for galaxy 04590. In a separate study, \cite{tacchella22} used the spectra from \cite{curti22} and NIRCam photometry with the \texttt{Prospector} code \citep{johnson21} to infer the gas-phase metallicities of objects 04590, 06355, and 10612. Interestingly, despite masking the [OIII]$\lambda$4363 line in their fits, they found broad agreement with the results of \cite{curti22}.

 \cite{trump22} calculated metallicities for all five $z>5$ galaxies starting from the level 2 2D spectral data products from the JWST Science Calibration Pipeline (version 1.5.3). They flux calibrated these spectra using simulations drawn from the NIRSpec Instrument Performance Simulator \citep{piqueras10}. Notably, \cite{trump22} found that the default 8 spatial pixel ``extended" pipeline spectral extraction aperture was too large for these compact objects, instead preferring a 4 spatial pixel extraction aperture.
 
\cite{trump22} then used the IDL function \texttt{mpfit} to fit Gaussian profiles to the individual emission lines. Using the direct $T_e$ method following \citet{nicholls20} and \citet{perez-montero21} and assuming a fixed Balmer decrement (see Section~\ref{metallicity} below), they reported high values of $T_e$ for both 04590 (22400~K) and 08140 (26900~K). They were the first to report [OIII]$\lambda$4363 detections and metallicities for objects 05144 and 08140.

\cite{rhoads22} released their own study of the metallicities of objects 04590, 06355, and 10612. They used the level 3 1D spectra provided by the JWST Science Calibration Pipeline (version 1.5.3). They fit Gaussian profiles to the individual emission lines for each observation (s007 and s008) separately, before averaging (with weighting) the resulting line ratios from each observation. They stated that their measured Balmer line ratios were within 1-2$\sigma$ of the theoretical values \citep{osterbrock89}, and they did not mention any flux recalibrations. They used the direct $T_e$ method detailed in \cite{jiang19}, which was based on \cite{izotov06}. They reported a very high $T_e$ (37000~K) for galaxy 04590, even exceeding those reported by \cite{curti22} and \cite{trump22}.

Given the broad range of methodologies, software, calibrations, and corrections used in the literature to study these five galaxies, our goals in this work are to summarize these methods and results (see Tables~\ref{tab:methods} and \ref{tab:literature}) and to perform our own measurements using improved methodologies that are less sensitive to calibration and flux extraction systematics. Additionally, we will compare our metallicities with those of a sample of extreme emission-line galaxies at $z\lesssim1$ from \cite{laseter22} as a function of the measured rest-frame equivalent width (EW) of the H$\beta$ line.

\section{Observations}\label{observations}
We analyze the JWST ERO of SMACS J0723 from Program \#2736,
focusing on the NIRSpec observations of objects 04590, 05144, 06355, 08140, and 10612---the five galaxies with spectroscopic redshifts $z>5$ from \cite{carnall22}. These observations were taken using the NIRSpec microshutter array in two pointings (s007 and s008) using both the G235M/F170LP and G395M/F290LP grating/filter combinations. Each combination of pointings and gratings/filters were observed for 8754 seconds. Specifically, we analyze the G395M/F290LP observations for objects 04950, 05144, 06355, and 10612, and the G235M/F170LP observations for object 08140. We use the level 3 1D and 2D spectral data products from the JWST Science Calibration Pipeline (version 1.5.3).

\section{Analysis}\label{analysis}
\subsection{Emission Line Measurement and Spectral Calibration}\label{sec:lines}

In our initial examination of the 1D (\texttt{x1d}) spectra, we found the same microshutter (for object 04590) and contamination problems previously noted in the literature. To minimize these and avoid any unwanted effects resulting from differences in total flux calibration between the two observations for each object, we masked areas affected by large, non-physical spikes in the spectral flux greater than the peak of the [OIII]$\lambda$5007 line before coadding the two spectra. We propagated the masked regions of each spectrum to the combined spectrum, preferring to exclude the union of these masked regions rather than introduce problems from unmatched flux calibrations between the two observations. Before measuring line fluxes, we converted the summed spectra from units of $f_{\nu}$ to units of $f_{\lambda}$. 

\begin{figure*}[ht]
\centering
\includegraphics[angle=0,width=0.49\textwidth]{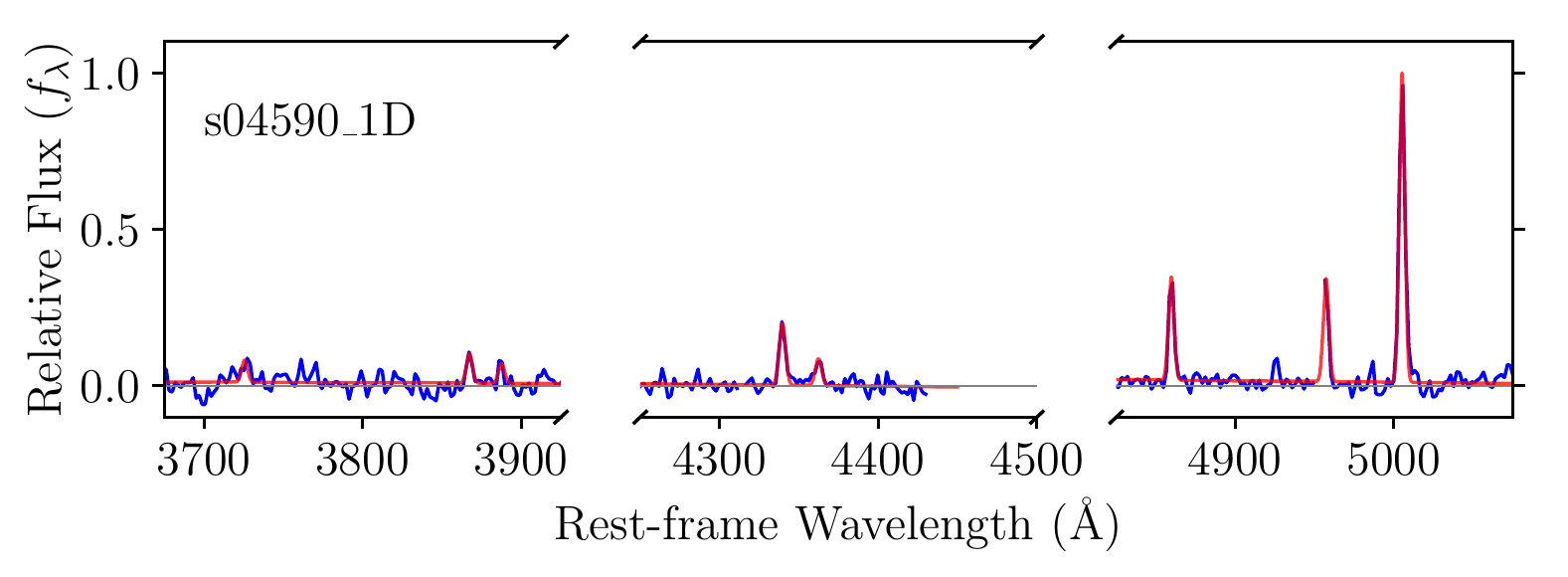}
\includegraphics[angle=0,width=0.49\textwidth]{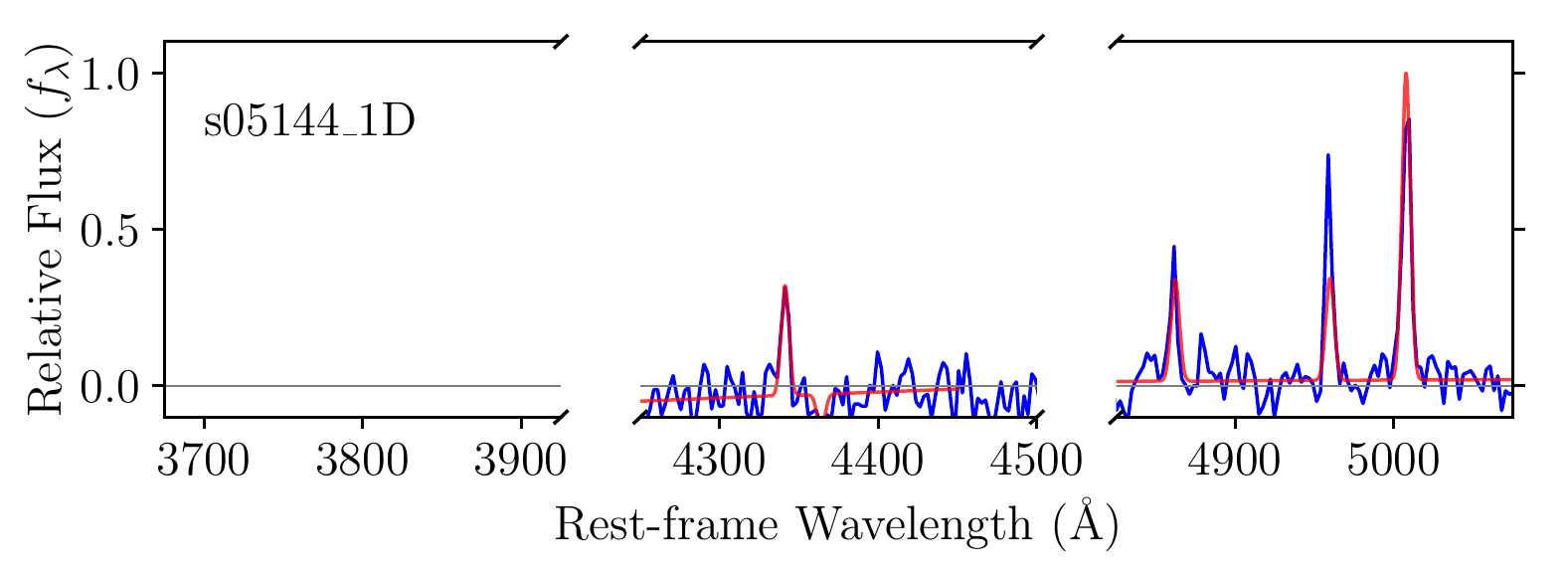}
\includegraphics[angle=0,width=0.49\textwidth]{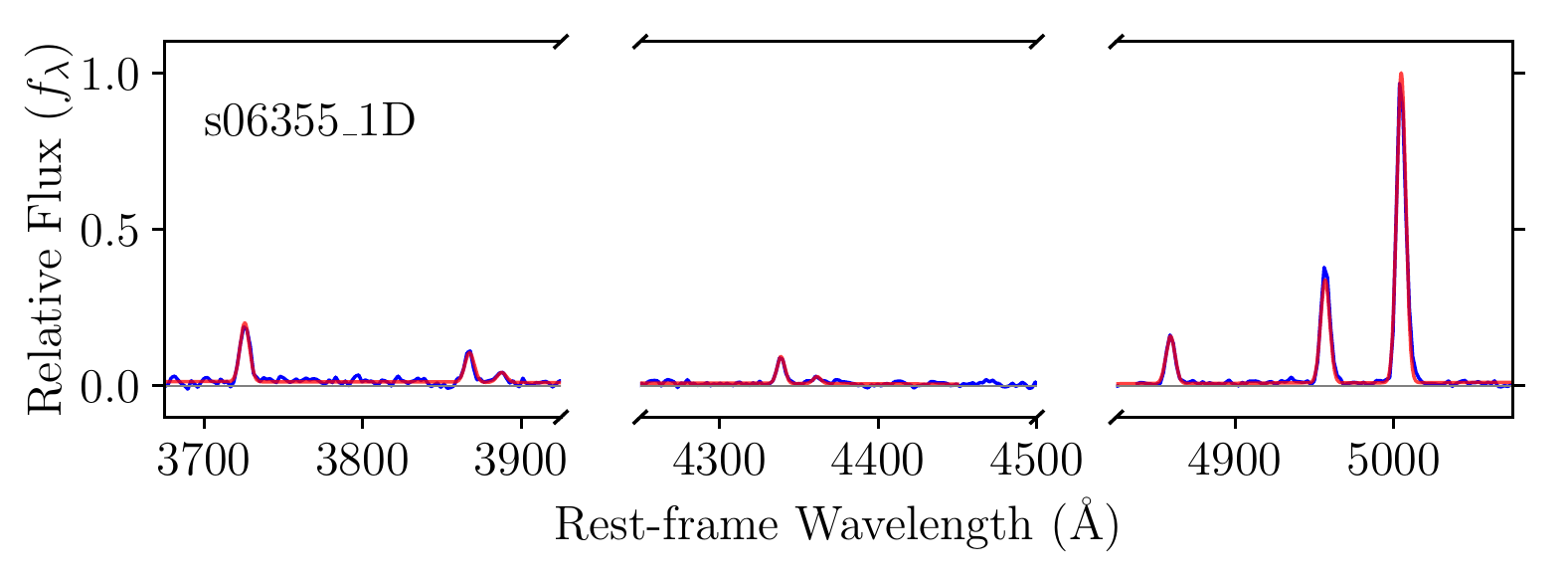}
\includegraphics[angle=0,width=0.49\textwidth]{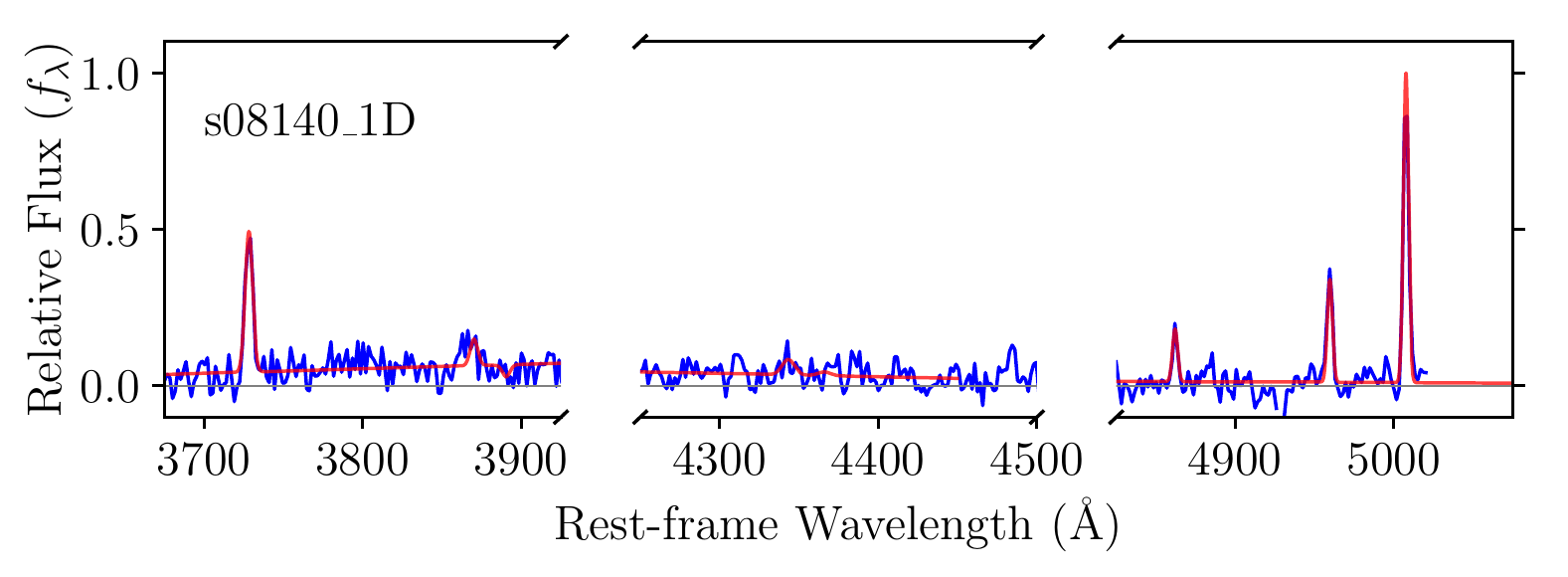}
\includegraphics[angle=0,width=0.49\textwidth]{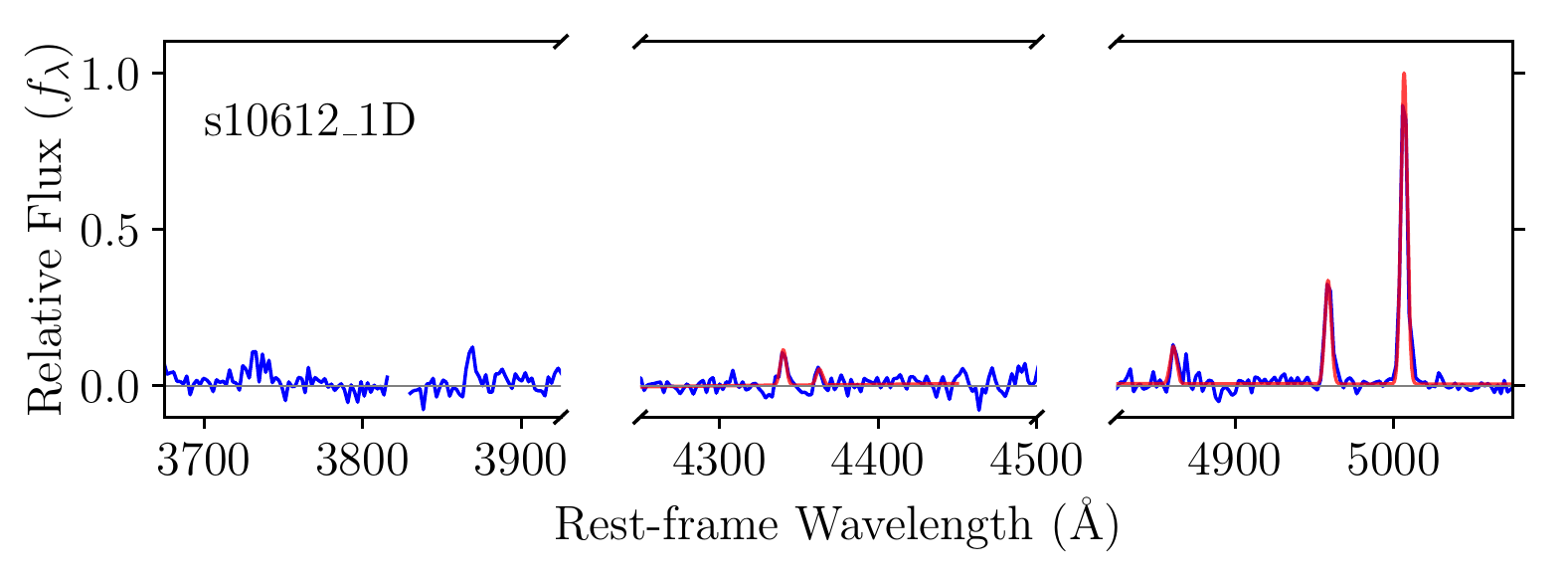}
\caption{Gaussian and continuum fits to the 1D spectra for the galaxy sample. In each panel, the blue curve is the coadded 1D spectrum, and the red curves are the best fit functions to the spectrum. For visual clarity, we normalize the flux scale to be equal to 1 at the peak of the [OIII]$\lambda$5007 line fit. Note that object 05144 is at too low of redshift for the G395/F290LP observations to capture the [OII]$\lambda$3727+H8+[NeIII]$\lambda$3869 complex. To avoid introducing errors from comparing lines across both filter/grating observations, we omit the [OII]$\lambda$3727+H8+[NeIII]$\lambda$3869 fit for this object. Also note that due to limited signal, the [OII]$\lambda$3727+H8+[NeIII]$\lambda$3869 fit for object 10612 failed to converge.}
\label{fig:1Dfits}
\end{figure*}

We measured emission-line fluxes in three groups: [OIII]$\lambda\lambda$5007,4959+H$\beta$, [OIII]$\lambda$4363+H$\gamma$, and [OII]$\lambda$3727+H8+[NeIII]$\lambda$3869. For each group of lines, we simultaneously fit Gaussian functions and a linear continuum (in $f_{\lambda}$) to the spectrum using the \texttt{scipy} function \texttt{curve\_fit} \citep{scipy}. When fitting [OIII]$\lambda\lambda$5007,4959+H$\beta$, we enforced a common line width for all three lines, as well as fixed ratios of the line centers and a fixed 3:1 ratio of the [OIII]$\lambda$5007 to [OIII]$\lambda$4959 line ratios. Similarly, when fitting [OIII]$\lambda$4363+H$\gamma$ and [OII]$\lambda$3727+H8+[NeIII]$\lambda$3869, we required a common line width for the lines and a fixed ratio of the line centers for each group. This procedure is helpful for compensating for regions of the [OIII]$\lambda\lambda$5007,4959 doublet that are masked due to contamination (for example, in 04590), as well as for better fitting relatively faint lines, such as [OIII]$\lambda$4363, by allowing the more strongly detected H$\gamma$ line to influence the line center and line width fits. Fitting faint lines with single Gaussians generally results in upward biasing.
We show the results of these emission-line fits in Figure~\ref{fig:1Dfits}.

Visually, it is clear that objects 04590, 06355, and 10612 are reasonably well fit by the Gaussian functions and fitting constraints, while the lack of any meaningful [OIII]$\lambda$4363 detection and overall lower signal-to-noise (S/N) in objects 05144 and 08140 result in unusable fits. This is consistent with 05144 and 08140 only being analyzed after the data reprocessing in \cite{trump22}, and being excluded from the other recent works. 

Due to these non-detections and the overall high noise levels in 05144 and 08140, we now turn to the 2D spectra (\texttt{s2d}). While \cite{curti22} and \cite{trump22} elected to recalibrate the level 2 2D spectra, our methods do not require a robust absolute flux calibration. Rather, we require only a firm relative flux calibration over short ($<$150~\AA{} rest-frame) spans in wavelength. As a result, the lack of flux calibrations in the level~3 2D spectra are inconsequential to our study. To prepare these data for emission-line measurements, we first coadded the two observations for each object. As with the 1D spectra, before measuring line fluxes, we converted from units of $f_{\nu}$ to units of $f_{\lambda}$. From these coadded 2D spectra, we performed emission-line fitting and extraction using two different methods.

First, we used a weighted spectral extraction to extract 1D spectra from the now coadded 2D spectra. While the level~3 1D spectra and \cite{trump22} used fixed extraction apertures (of 8 pixels and 4 pixels, respectively), we instead used a weighting scheme. For each object, we examined the region around the [OIII]$\lambda$5007 line---the highest signal emission feature---in a 20~\AA{} window. We summed the flux in this region over the spectral direction to construct a 1D spatial line profile. We renormalized this line profile so its peak value was 1, making it usable as a spatial weighting function. We then applied this weighting function to the 2D spectrum and summed over the spatial axis to produce a 1D spectrum. Finally, we fit the emission lines using the same methodology that we used on the original 1D spectra.

\begin{figure*}[ht]
\centering
\includegraphics[angle=0,width=0.49\textwidth]{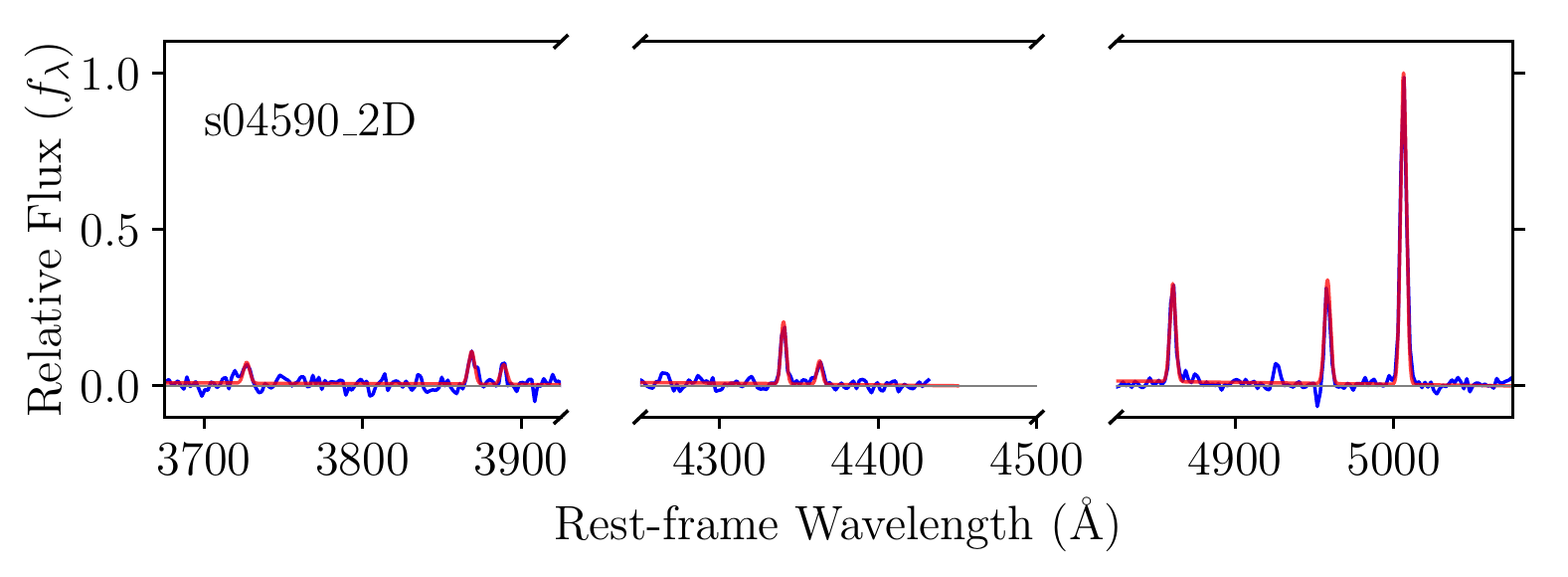}
\includegraphics[angle=0,width=0.49\textwidth]{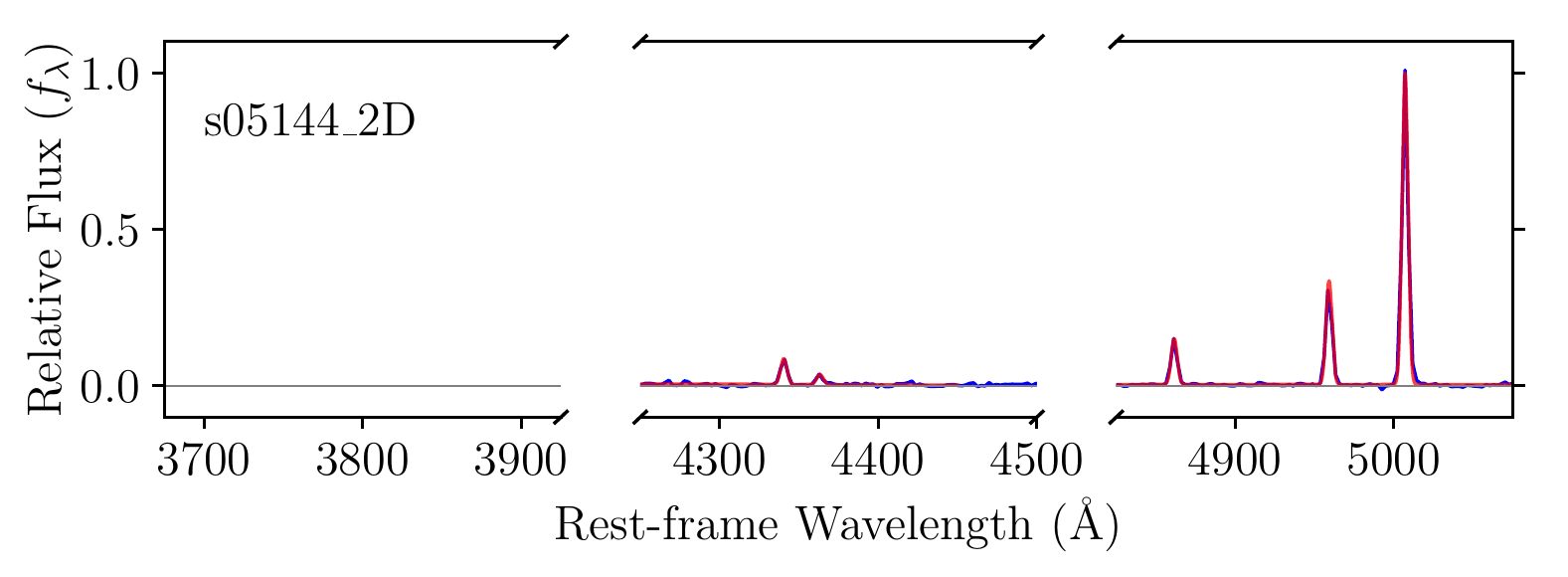}
\includegraphics[angle=0,width=0.49\textwidth]{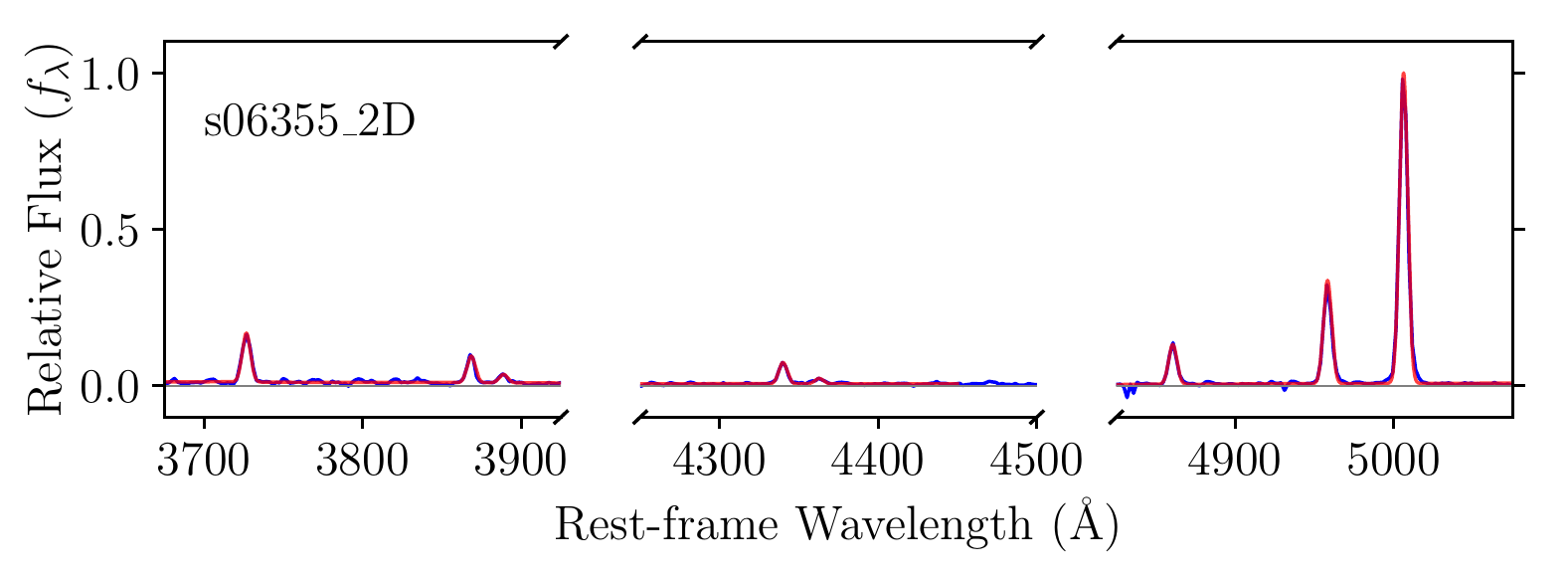}
\includegraphics[angle=0,width=0.49\textwidth]{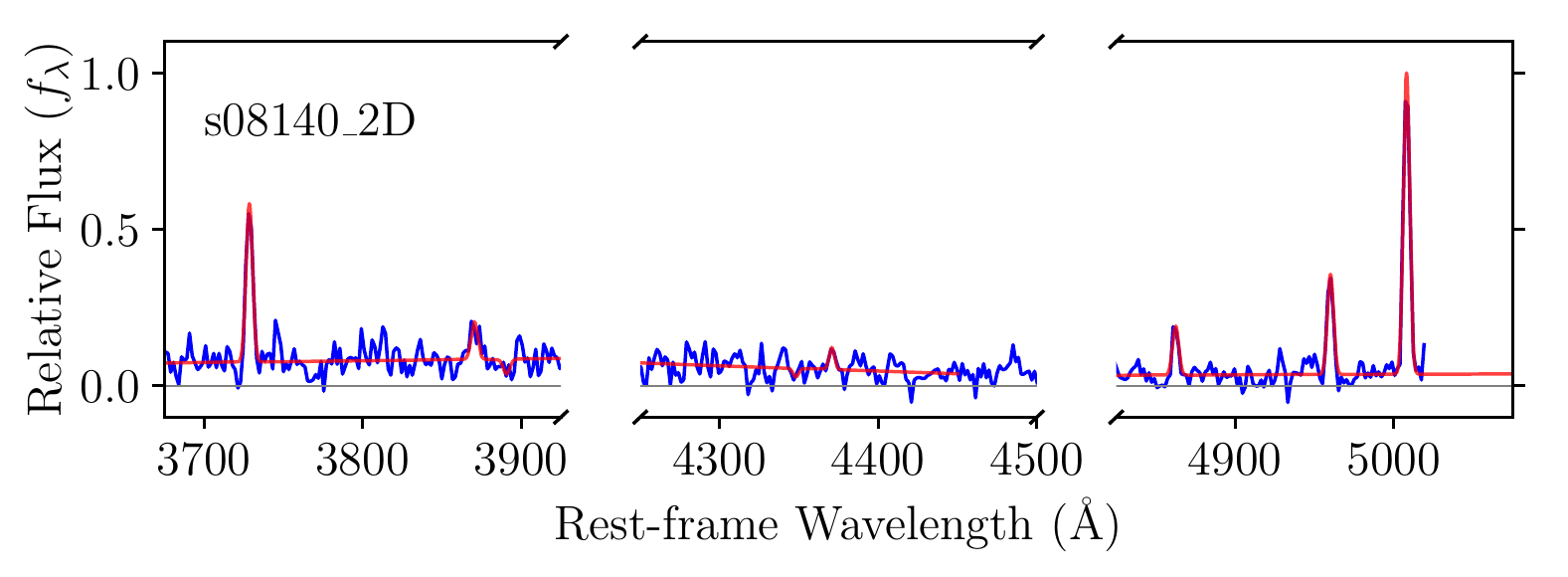}
\includegraphics[angle=0,width=0.49\textwidth]{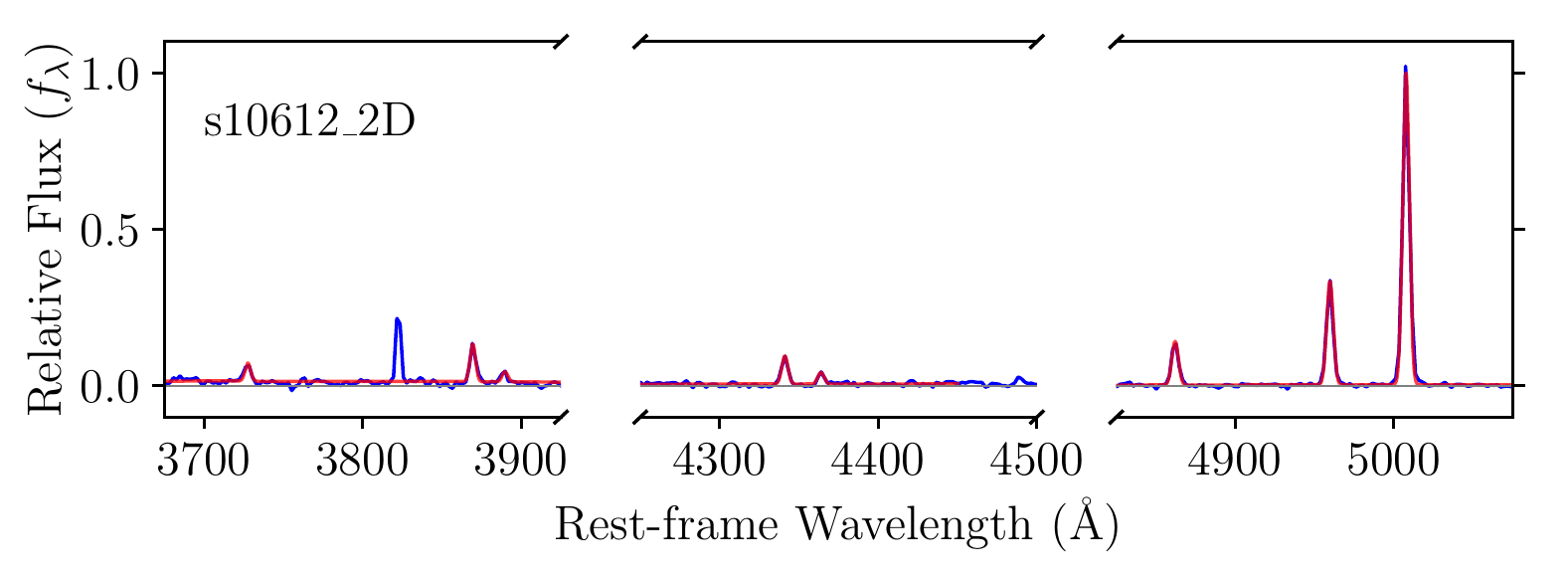}
\caption{Gaussian and continuum fits to the weighted aperture extracted 1D spectra for the galaxy sample. In each panel, the blue curve is the coadded, extracted 1D spectrum, and the red curves are the best fit functions to the spectrum. For visual clarity and ease of comparison, we again normalize the flux scale to be equal to 1 at the peak of the [OIII]$\lambda$5007 line fit.}
\label{fig:2Dfits}
\end{figure*}

We show our extracted 1D spectra and the corresponding emission-line fits in Figure~\ref{fig:2Dfits}. 
In all cases, the noise in the spectral continuum relative to the emission line strengths is greatly decreased for all 5 objects relative to the original 1D spectra. Objects 05144 and 10612 show more consistent [OIII]$\lambda\lambda$5007,4959 internal line ratios, and the contamination around the H$\beta$ line in 10612 is completely eliminated. Most notably, the new spectrum for object 05144 shows clear and well-fit [OIII]$\lambda$4363 emission where the original 1D spectrum showed only noise. Only object 08140 fails to show any significant H$\gamma$ or [OIII]$\lambda$4363 emission above the continuum noise level. This weighted extraction removes any systematics introduced in the choice of a fixed extraction aperture. The weighted scheme best captures the line and continuum flux of an object based on its individual spatial point-spread characteristics for extraction into a 1D spectrum. 

In an attempt to further increase the signal in object 08140, we implemented one final flux measurement technique. Here we again used our coadded 2D spectra. However, instead of using weighted spatial apertures to produce a 1D spectrum for line flux measurements, we performed direct line flux extraction in the 2D spectra. 

\begin{figure*}[ht]
\centering
\includegraphics[angle=0,width=0.19\textwidth]{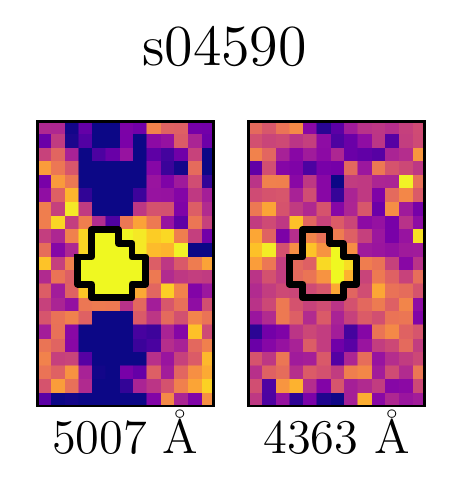}
\includegraphics[angle=0,width=0.19\textwidth]{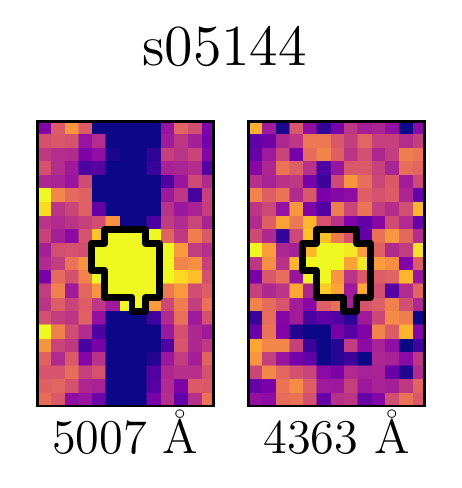}
\includegraphics[angle=0,width=0.19\textwidth]{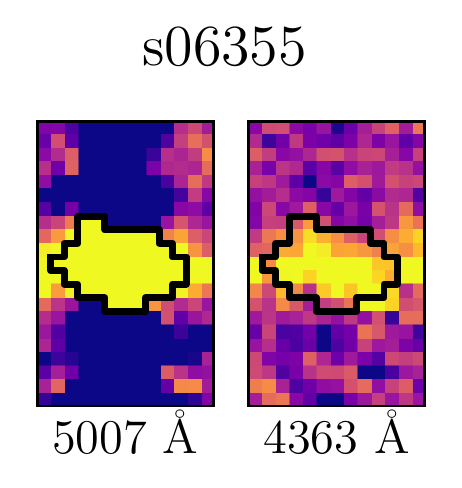}
\includegraphics[angle=0,width=0.19\textwidth]{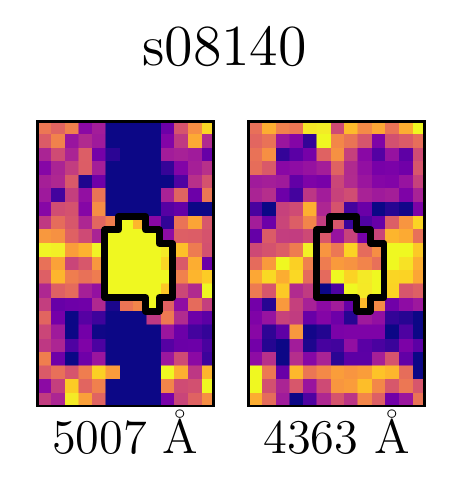}
\includegraphics[angle=0,width=0.19\textwidth]{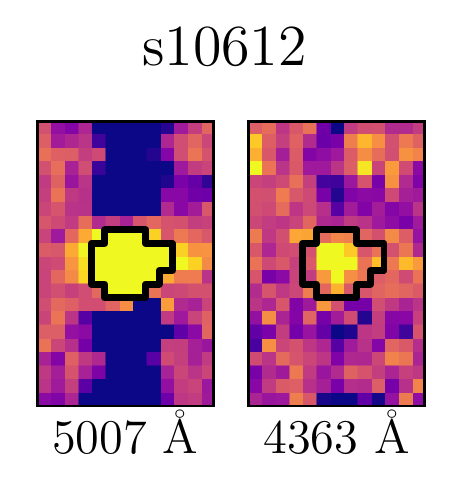}
\caption{Cutouts of the coadded 2D spectra centered on the [OIII]$\lambda$5007 and [OIII]$\lambda$4363 lines. The extraction footprints determined by the [OIII]$\lambda$5007 line shape are shown with black outlines.}
\label{fig:footprints}
\end{figure*}

We first analyzed the [OIII]$\lambda$5007 line and fit its 2D footprint in the 2D spectrum by eye. We summed the flux in this footprint to measure the [OIII]$\lambda$5007 line flux. We then translated this footprint to the positions of the other emission lines of interest. At each position, we again summed the flux within the footprint to measure that line's flux. To remove the underlying continuum at each line measurement, we used the same footprint to sample the continuum at random positions near the emission lines of interest, and we subtracted the median of these measurements from each line flux. We show these [OIII]$\lambda$5007 extraction footprints in Figure~\ref{fig:footprints}. 

Using this method, we detected faint but measurable [OIII]$\lambda$4363 flux in object 08140, sufficient for calculating its metallicity. We consider this flux measurement method to be superior to 1D spectral fitting and 2D spectral weighted extraction, as it inherently removes any systematic effects introduced by noise or contamination along the spatial axis. Directly measuring the line fluxes in the 2D spectra based on the shape of the highest S/N line better captures the emission signal of both strong and faint lines by better excluding nearby non-emission pixels that would otherwise introduce additional noise or continuum contributions to the measurements.

To estimate the uncertainties on these line flux measurements, we performed Monte Carlo simulations for each object and methodology. For a given 1D or 2D spectrum, we constructed an array of Gaussian distributions centered at the flux values of the spectra with standard deviations given by the flux errors provided in the 1D and 2D data files. We sampled these distributions 5000 times for each object and method and re-extracted the line fluxes from the resulting spectra. We then took the 16th and 84th percentile of each of the resulting distributions as the $1\sigma$ confidence interval for the line flux and EW measurements.

We compare the performance and fidelity of all of these techniques in Section~\ref{sec:results}.

\begin{deluxetable*}{rccccc}[ht]
\renewcommand\baselinestretch{1.0}
\tablewidth{0pt}
\tablecaption{Methods}
\label{tab:methods}
\tablehead{Study & Initial Data & Calibration & Line Measurement & Balmer Correction & Metallicity Prescription}
\startdata
 \multirow{2}{*}{This work: 1D} & Level 3 & \multirow{2}{*}{JWST Pipeline} & Simultaneous Gaussian & Fixed & \multirow{2}{*}{\cite{izotov06}} \\ 
 & 1D Spectra  &  &  line \& linear continuum fits & H$\gamma$/H$\beta$=0.47 &  \\ \hline
 \multirow{3}{*}{This work: 2D} & \multirow{2}{*}{Level 3} & \multirow{3}{*}{JWST Pipeline} & Weighted Extraction & \multirow{2}{*}{Fixed} &\multirow{3}{*}{\cite{izotov06}} \\ 
 & \multirow{2}{*}{2D Spectra}  &  & Simultaneous Gaussian  & \multirow{2}{*}{H$\gamma$/H$\beta$=0.47} & \ \\
 &  &  & line \& linear continuum fits &  & \\ \hline
 \multirow{2}{*}{This work: Footprint} & Level 3 & \multirow{2}{*}{JWST Pipeline} & \multirow{2}{*}{2D Footprints} & Fixed & \multirow{2}{*}{\cite{izotov06}} \\ 
 & 2D Spectra  &  &  & H$\gamma$/H$\beta$=0.47 &  \\ \hline
\multirow{2}{*}{\cite{schaerer22}} & Level 3 & \multirow{2}{*}{JWST Pipeline} & Gaussian line \& & \multirow{2}{*}{Power law fit} & \multirow{2}{*}{\cite{izotov06}} \\
 & 1D Spectra &  & flat continuum fits &  & \\ \hline
\multirow{2}{*}{\cite{curti22}} & Level 2 2D  & \multirow{2}{*}{GTO Pipeline} & \multirow{2}{*}{\texttt{PPXF} fitting} & \multirow{2}{*}{None} & \cite{nicholls13}, \\
 & count-rate maps & & & & \texttt{PYNEB} \\ \hline
\multirow{2}{*}{\cite{trump22}} & Level 2 2D  & \multirow{2}{*}{NIRSpec Simulator} & 4 pixel 2D extraction  & Fixed & \cite{nicholls20},\\ 
 & count-rate maps & & Gaussian line fits & H$\gamma$/H$\beta$=0.47 & \cite{perez-montero21} \\ \hline
\multirow{2}{*}{\cite{rhoads22}} & Level 3 & \multirow{2}{*}{JWST Pipeline} & Simultaneous Gaussian  & \multirow{2}{*}{None} & \cite{jiang19}, \\ 
 & 1D Spectra  &  & line \& flat continuum fits &  & \cite{izotov06} \\ \hline
\enddata
\end{deluxetable*}

\subsection{Metallicity Calculations}\label{metallicity}
To calculate the metallicities for all our methodologies, 
we followed \cite{laseter22}, who used the direct $T_e$ method described in \cite{izotov06} (which was also used by \citealt{schaerer22} and \citealt{rhoads22}). This means we will be able to make direct comparisons with \cite{laseter22}'s $z\lesssim1$ extreme emission-line galaxy sample in Section~\ref{sec:laseter}.

In this method, the line ratios of [OIII]$\lambda\lambda$5007,4959 and [OIII]$\lambda$4363 are used to derive an [OIII] $T_e$. This $T_e$ may then be used with the [OIII]$\lambda\lambda$5007,4959 to H$\beta$ and [OII]$\lambda$3727 to H$\beta$ line ratios to derive the abundances of O$^{+}$/H and O$^{++}$/H (see Equations 1, 2, 3, and 5 in \citealt{izotov06}). 

Again following \cite{laseter22}, we adopted a modification to this method (also used by \citealt{trump22}), where instead of measuring the ratio of [OIII]$\lambda\lambda$5007,4959 to [OIII]$\lambda$4363 directly, we instead used the following equation:
\begin{equation}
 \frac{\textrm{[OIII]}_{5007,4959}}{\textrm{[OIII]}_{4363}} = \frac{\textrm{[OIII]}_{5007,4959}}{\textrm{H}\beta} \times \frac{\textrm{H}\gamma}{\textrm{[OIII]}_{4363}} \times 0.47 \,.
 \end{equation}
This equation relates the [OIII] lines to nearby Hydrogen lines and enforces an H$\gamma$/H$\beta$ Balmer decrement equal to the extinction-free Case B recombination value of 0.47 \citep{osterbrock89}. This helps to correct for any relative miscalibration of the observed spectral flux as a function of wavelength by considering ratios of nearby ($<$150~\AA{} rest-frame) lines before using the Balmer decrement to compare these flux ratios across the much larger $\sim640$~\AA{} (rest-frame) span. It also implicitly corrects the line ratios for any dust extinction. 

Similarly, to ensure a proper ratio of [OII]$\lambda$3727 to H$\beta$, we use the H8/H$\beta$ Case B recombination value of 0.107 \citep{osterbrock89}:
\begin{equation}
\frac{\textrm{[OII]}_{3727}}{\textrm{H}\beta} = \frac{\textrm{[OII]}_{3727}}{\textrm{H}8} \times 0.107 \,.
\end{equation}

To estimate the uncertainties on our measurements of $T_e$ and metallicity, we extend the Monte Carlo simulations described in Section~\ref{sec:lines} to recalculate the $T_e$ and metallicity for each Monte Carlo run. We again take the 16th and 84th percentile of each of the resulting distributions of $T_e$ and metallicity as the $1\sigma$ confidence interval for the measurements.

\section{Results}\label{sec:results}

\begin{deluxetable*}{rccccc}[ht]
\renewcommand\baselinestretch{1.0}
\tablewidth{0pt}
\tablecaption{Metallicities and Electron Temperatures}
\label{tab:literature}
\tablehead{Object & 04590 & 05144 & 06355 & 08140 & 10612}
\startdata
Redshift ($z$) & 8.4989 & 6.3805 & 7.6687 & 5.2753 & 7.6607  \\ \hline
1D 12+log(O/H) & $6.84^{+0.08}_{-0.07}$ & \nodata & $7.89^{+0.07}_{-0.07}$ & \nodata & $7.82^{+0.15}_{-0.11}$ \\
2D 12+log(O/H) & $6.98^{+0.97}_{-0.15}$ & $7.69^{+0.28}_{-0.17}$ & $8.06^{+0.20}_{-0.11}$ & \nodata & $7.72^{+0.31}_{-0.10}$ \\
Footprint 12+log(O/H) & $6.82^{+0.29}_{-0.10}$ & $7.63^{+0.22}_{-0.23}$ & $8.03^{+0.19}_{-0.17}$ & $7.68^{+0.51}_{-0.50}$ & $7.67^{+0.21}_{-0.21}$ \\ 
\cite{schaerer22} 12+log(O/H) & \nodata & \nodata & 7.85 & \nodata & 7.85 \\
\cite{curti22} 12+log(O/H) & 6.99$\pm$0.11 & \nodata & 8.24$\pm$0.07 & \nodata & 7.73$\pm$0.12 \\
\cite{trump22} 12+log(O/H) & $<$7.75 & 7.98 & 8.18 & $<$7.75 & 7.95 \\
\cite{rhoads22} 12+log(O/H) & 6.88$\pm$0.15 & \nodata & 8.09$\pm$0.16 & \nodata & 7.68$\pm$0.24 \\ \hline
1D $T_e$ ($10^4$ K) & $3.38^{+0.35}_{-0.34}$ & \nodata & $1.49^{+0.09}_{-0.09}$ & \nodata & $1.72^{+0.20}_{-0.20}$ \\
2D $T_e$ ($10^4$ K) & $2.87^{+0.82}_{-1.15}$ & $1.79^{+0.27}_{-0.31}$ & $1.37^{+0.12}_{-0.16}$ & \nodata & $1.79^{+0.17}_{-0.33}$ \\
Footprint $T_e$ ($10^4$ K)  & $3.48^{+0.42}_{-0.96}$ & $1.83^{+0.37}_{-0.25}$ & $1.44^{+0.17}_{-0.16}$ & $1.77^{+0.14}_{-1.77}$ & $1.83^{+0.34}_{-0.25}$ \\
\cite{schaerer22} $T_e$ ($10^4$ K) & \nodata & \nodata & 1.60 & \nodata & 1.87 \\
\cite{curti22} $T_e$ ($10^4$ K) & 2.77$\pm$0.42 & \nodata & 1.20$\pm$0.07 & \nodata & 1.75$\pm$0.16 \\
\cite{trump22} $T_e$ ($10^4$ K) & 2.24 & 1.58 & 1.29 & 2.69 & 1.58 \\
\cite{rhoads22} $T_e$ ($10^4$ K) & 3.72$\pm$0.99 & \nodata & 1.34$\pm$0.16 & \nodata & 2.19$\pm$0.54 \\ 
\enddata
\tablecomments{We convert the $Z/Z_{\odot}$ values for metallicity given in \cite{trump22} to 12+log(O/H) assuming a solar abundance of 8.75 \citep{bergemann21}.}
\end{deluxetable*}

The metallicity results from this work and from the literature demonstrate reasonable agreement across all five objects (see Table~\ref{tab:literature}). Our 1D spectral results most closely resemble the results of \cite{schaerer22}, who also used the level~3 1D spectra.
Interestingly, \cite{rhoads22} used the same 1D spectra, but their results show values more similar to our 2D methods.

Our 2D and footprint methods show strong agreement with one another and most closely resemble the results of \cite{rhoads22}. The results of \cite{trump22} seem to be consistently $\sim$0.2~dex higher than our own. Interestingly, \cite{trump22} reported using a Balmer decrement correction identical to our own (see their Equation~1 and our Equation~1), so we attribute the differences in calculated metallicity to their use of the \cite{perez-montero21} metallicity estimator instead of that of \cite{izotov06}; the former has a flatter slope in metallicity as a function of $T_e$ at  $T_e>15\,000\textrm{~K}$.

The results from \cite{curti22} are also $\sim$0.1~dex higher than our 2D methods, but they show general agreement within the error bounds of both studies. Especially given its performance in detecting [OIII]$\lambda$4363 in object 08140, we believe that our footprint extraction method is the best method for line flux measurements, as it helps to avoid any aperture effects, as well as noise along the spatial axis and any effects from the ``nods" in the 2D data.

For all of the objects, we find that the contributions to the overall 12+log(O/H) abundance from O$^+$ are $\lesssim$10\% of the contribution from O$^{++}$, indicating that these objects are all highly ionized. 

For $T_e$, all of the studies seem to agree with one other within their error bounds. We note---echoing the remarks from the literature---that all values of $T_e$ calculated for object 04590 are unphysically high, or at the upper bound of reasonability. Of all the studies, the recalibrated spectra from \cite{curti22} and \cite{trump22} give the lowest values. The large uncertainty in our 2D results for the $T_e$ for object 04590 (and the corresponding large uncertainty in metallicity) also reflect the relatively poor data quality in the 04590 spectra. Due to these uncertain measurements of $T_e$ (the result of extraordinarily bright [OIII]$\lambda$4363 emission relative to [OIII]$\lambda\lambda$5007,4959), the galaxy's high redshift of $z=8.4989$ \citep{schaerer22}, and the partial data contamination from a malfunctioning microshutter, 04590 may warrant additional JWST or other multiwavelength followup. 

\subsection{Metallicity and H$\beta$ Equivalent Width} \label{sec:laseter}

\cite{laseter22} presented a correlation between direct $T_e$ metallicity and rest-frame EW(H$\beta$) at rest-frame EW(H$\beta$)$>100$~\AA{}(Spearman rank coefficient $=-0.469$, $p=0.037$) using a sample of emission-line galaxies at $z\lesssim1$ observed with the DEIMOS multiobject spectrograph on the Keck~II 10~m telescope. They proposed that galaxies with high rest-frame EW(H$\beta$) may serve as analogs to some of the earliest galaxies. Given the high redshifts of the present sample, we now make comparisons with the \cite{laseter22} sample. 

We measured the EWs for the 1D and 2D weighted aperture extracted spectra through the simultaneous fitting of the H$\beta$ (and [OIII]$\lambda\lambda$5007,4959) line fluxes and the nearby continuum flux density level. Since we shifted the wavelength scale to the rest frame in the first step of our fitting, dividing the integrated line flux by the continuum flux density provides the proper rest-frame line EW. 

\begin{figure*}[ht]
\centering
\includegraphics[angle=0,width=\textwidth]{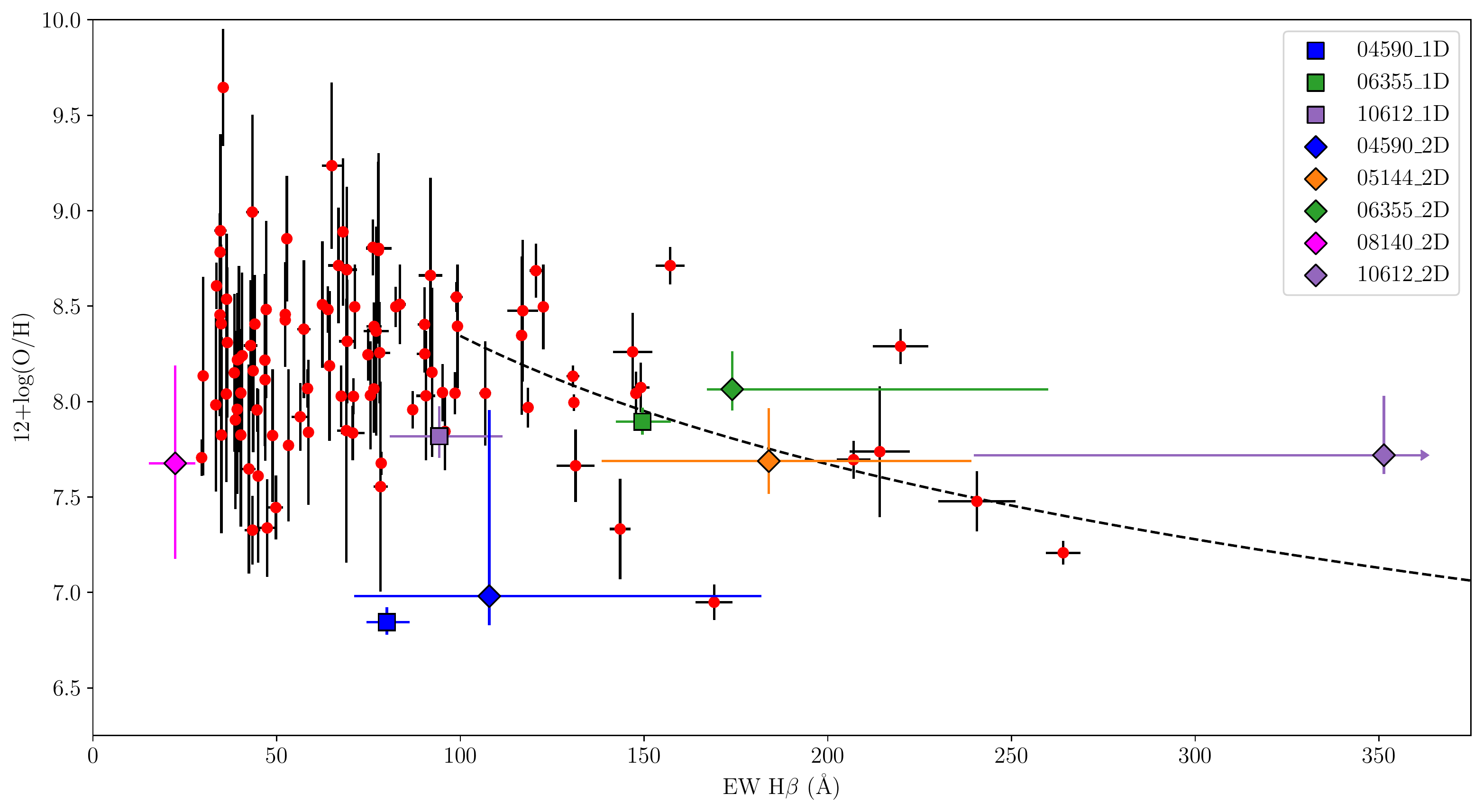}
\caption{Metallicity vs. rest-frame EW(H$\beta$) for the \cite{laseter22} sample of galaxies at $z\lesssim1$ with strong emission lines and rest-frame EW(H$\beta)>30$~\AA\  (red circles), JWST 1D spectra results (squares, see legend for colors), and JWST 2D spectra results (diamonds, see legend for colors). Note that the 2D spectrum for object 05144 has very faint continuum near the H$\beta$ line. Given the substantial error on the continuum for this object, the upper error bounds on the EW(H$\beta$) exceed $\sim$400~\AA{}, so we use a rightward-pointing arrow. Note that as object 08140 only showed an [OIII]$\lambda$4363 detection with the footprint method, we use that metallicity with the 2D EW(H$\beta$) (pink diamond). The black dashed curve is the best fit to 12+log(O/H) vs. EW(H$\beta$) for values of EW(H$\beta)>100$~\AA{} from \cite{laseter22}.
}
\label{fig:laseter}
\end{figure*}

Objects 04590, 06355 and 08140 show the strongest continuum detections in both their 1D and 2D spectra, which is reflected in their lower EWs and in their relatively smaller error bounds (see Figure~\ref{fig:laseter}). For object 10612, the 2D spectrum shows very weak continuum emission. As a result, 10612 exhibits a very high  ($>$300~\AA{}) 2D EW(H$\beta$) with very large error bounds due to the high fractional flux error in the continuum measurement. Due to the fixed size 2D to 1D flux extraction apertures from the JWST Pipeline and the lack of 1D [OIII]$\lambda$4363 detections for objects 05144 and 08140, we consider the 2D EWs---despite their large error bounds---to be the better (if less certain) measurements of EW(H$\beta$) for all five objects. Moreover, we agree with the claims of \cite{trump22}, \cite{rhoads22}, and \cite{brinchmann22} that the uncertainties provided for the 1D spectra underestimate the true uncertainties by a factor of $\sim$2. Thus, our calculated uncertainties for the 1D EW(H$\beta$) are likely also underestimated.

We compare our results with the \cite{laseter22} sample in Figure~\ref{fig:laseter}, finding good agreement. The objects with 2D EW(H$\beta$)$>100$~\AA{} (04590, 05144, 06355, and 10612) deviate from the \cite{laseter22} best-fit 12+log(O/H) curve by $-1.29$, $-0.06$, 0.26, and 0.59 dex, respectively. Given the before-mentioned high $T_e$ for object 04950 and the large uncertainty in EW(H$\beta$) for object 10612, the sample shows reasonable agreement with the \cite{laseter22} curve. This suggests that low-redshift, high EW(H$\beta$) objects may indeed be good analogs of early galaxies at $z\sim5-9$.

\cite{laseter22} also fit a metallicity--EW(H$\beta$)--age relation in their study for galaxies with EW(H$\beta>100$~\AA{}), assuming a continuous starburst model. Using their relation of $\log{t_{\textrm{years}}}=12+\log{\textrm{O/H}}-0.93$ and our best footprint method metallicities, we estimate galaxy ages of 04590: $0.8^{+0.7}_{-0.2}$, 05144: $5.1^{+3.5}_{-2.1}$, 06355: $12.6^{+6.9}_{-4.1}$, 08140: $5.6^{+12.6}_{-3.4}$, and 10612: $5.5^{+3.5}_{-2.1}$~Myr. Even 12.6~Myr for object 06355---the oldest estimated galaxy age in the sample---is reasonable, as the age of the universe was $\sim$660~Myr at $z=7.6687$ (assuming a flat cosmology with $\Omega_{\rm M}$=0.3, $\Omega_{\Lambda}$=0.7, and H$_0$=70~km~s$^{-1}$~Mpc$^{-1}$). 

\cite{carnall22} used \texttt{Bagpipes} \citep{carnall18} to estimate mean stellar ages of 04950: $2^{+10}_{-1}$, 05144: $1.2^{+0.3}_{-0.2}$, 06355: $1.2^{+0.3}_{-0.2}$, 08140: $19^{+21}_{-10}$, and 10612: $1.2^{+0.3}_{-0.2}$~Myr. Using \texttt{Prospector} fitting, \cite{tacchella22} found half-mass times ($t_{50}$) of 04950: $5^{+102}_{-3}$, 06355: $3^{+29}_{-1}$, and 10612: $7^{+96}_{-4}$~Myr. While these different age metrics are not directly comparable with one another, our estimates show reasonable agreement with both \cite{carnall22} and \cite{tacchella22} within the stated uncertainties and differences in spectral calibration.

\section{Summary}\label{sec:summary}
In this work, we measured gas phase metallicities for the five $z>5$ emission-line galaxies from the JWST ERO of SMACS 0723. 
We measured lines fluxes from these data using improved 1D and 2D spectral methods, which we consider to be optimal. 
We then computed metallicities using the direct $T_e$ method. We compared our results with those in the literature and found 
reasonably good agreement within the provided error bounds, which is encouraging 
given the variety of data processing and line ratio measurement methodologies utilized.

We found that the metallicities and EW(H$\beta$) of the $z>5$ galaxies roughly follow the trends of \cite{laseter22}'s
sample of $z\lesssim 1$ high rest-frame EW(H$\beta$) galaxies, offering support that they may be low-redshift analogs 
to the early galaxies seen with JWST. Finally, we estimated that the high-redshift galaxies are young compared to the age of the universe.

\acknowledgements

We thank the anonymous reviewer and data editor for their constructive report that helped us to improve this work.

We gratefully acknowledge the William F. Vilas Estate (A.J.T.)
and a Kellett Mid-Career Award and a WARF Named 
Professorship from the University of Wisconsin-Madison
Office of the Vice Chancellor for Research and Graduate Education
with funding from the Wisconsin Alumni Research Foundation (A.J.B.).

This work is based on observations made with the NASA/ESA/CSA James Webb Space Telescope (JWST). 
The data were obtained from the Mikulski Archive for Space Telescopes (MAST) at the Space Telescope 
Science Institute, which is operated by the Association of Universities for Research in 
Astronomy, Inc., under NASA contract NAS 5-03127 for JWST. 
These observations are associated with program \#2736. The specific observations analyzed 
can be obtained from MAST via \dataset[10.17909/espg-fy96]{https://doi.org/10.17909/espg-fy96}.

The Early Release Observations and associated materials were developed, executed, and compiled by the ERO production team:  Hannah Braun, Claire Blome, Matthew Brown, Margaret Carruthers, Dan Coe, Joseph DePasquale, Nestor Espinoza, Macarena Garcia Marin, Karl Gordon, Alaina Henry, Leah Hustak, Andi James, Ann Jenkins, Anton Koekemoer, Stephanie LaMassa, David Law, Alexandra Lockwood, Amaya Moro-Martin, Susan Mullally, Alyssa Pagan, Dani Player, Klaus Pontoppidan, Charles Proffitt, Christine Pulliam, Leah Ramsay, Swara Ravindranath, Neill Reid, Massimo Robberto, Elena Sabbi, Leonardo Ubeda. The EROs were also made possible by the foundational efforts and support from the JWST instruments, STScI planning and scheduling, and Data Management teams.


\facilities{JWST}

\software{astropy: \cite{astropy:2013, astropy:2018}}

\clearpage

\bibliography{ref1}

\end{document}